\documentclass[useAMS]{mn2e}
\usepackage{graphicx}
\usepackage{epsfig}

\def\ltsima{$\; \buildrel < \over \sim \;$}
\def\lsim{\lower.5ex\hbox{\ltsima}}
\def\gtsima{$\; \buildrel > \over \sim \;$}
\def\gsim{\lower.5ex\hbox{\gtsima}}

\newcommand{\be}{\begin{equation}}
\newcommand{\en}{\end{equation}}

\def\cmdue {\rm \ cm^{-2}}

\begin{document}
\title[The column densities of a complete sample of bright {\it Swift} GRBs ]{The X--ray absorbing column density of a complete sample of bright 
{\it Swift} Gamma--Ray Bursts} 

\author[S. Campana et al.]{S. Campana$^{1,}$\thanks{E-mail: sergio.campana@brera.inaf.it}, R. Salvaterra$^{2}$,
A. Melandri$^1$, S. D. Vergani$^{1}$,  S. Covino$^{1}$, 
\newauthor P. D'Avanzo$^{1}$, D. Fugazza$^{1}$, G. Ghisellini$^1$, B. Sbarufatti$^{1}$ \& G. Tagliaferri$^{1}$ \\
$^1$ INAF-Osservatorio Astronomico di Brera, Via Bianchi 46, I--23807 Merate (LC), Italy\\
$^2$ INAF-IASF Milano, Via Bassini 15, I--20133 Milano, Italy\\
%$^3$ Service d'Astrophysique, DSM/IRFU/SAp, CEA-Saclay, F--91191 Gif-sur-Yvette, France
%$^4$ SISSA, Via Bonomea 265, I--34136 Trieste, Italy
}

\maketitle

\begin{abstract}
A complete sample of bright {\it Swift} Gamma--ray Bursts (GRBs) has been recently selected by Salvaterra et al. (2011). 
The sample has a high level of completeness in redshift ({\bf 90}$\%$). We derive here the intrinsic absorbing X--ray column densities of these
GRBs making use of the {\it Swift} X--ray Telescope data. This distribution has a mean value of $\log(N_H/{\rm cm^{-2}})=21.7\pm0.5$.
This value is consistent with the distribution of the column densities derived from the total sample of GRBs with redshift. 
We find a mild increase of the intrinsic column density with redshift. This can be interpreted as due to the contribution 
of intervening systems along the line of sight.
Making use of the spectral index connecting optical and X--ray fluxes at 11 hr ($\beta_{OX}$), we investigate the relation of 
the intrinsic column density and the GRB `darkness'. We find that there is a very tight correlation between dark GRBs and high X--ray 
column densities. This clearly indicates that the dark GRBs are formed in a metal-rich environment where dust must be present.

\end{abstract}


\section{Introduction}

Long duration Gamma--ray bursts (GRBs) are thought to originate from the collapse of massive stars.
Several lines of evidence points toward this conclusion, ranging from the association to type Ib/c supernovae (Woosley 
\& Bloom 2006 and references therein), to the occurrence of GRBs in the most luminous part of their host galaxies 
(Svensson et al. 2010). The ambient medium in which GRBs explode is expected to be denser than the interstellar medium
and typical of star forming regions. 
The values of the absorbing column densities as measured in X--rays are high (Galama \& Wijers 2001;
Stratta et al. 2004; Campana et al. 2006; Watson et al. 2007).
An analysis of the intrinsic column densities of all {\it Swift} GRBs observed within 1,000 s and with a known redshift has been 
carried out by Campana et al. (2010). The selected sample consisting of 93 GRBs was biased. 
The distribution of the intrinsic X--ray absorption column density is consistent with a 
lognormal distribution with mean $\log N_H(z)=21.9\pm0.1$ ($90\%$ confidence level). 
This distribution is in agreement with the expected distribution for GRBs occurring randomly in giant molecular 
clouds similar to those within the Milky Way (Campana et al. 2006; Reichart \& Price 2002).
Looking at the distribution of X--ray column densities vs. redshift, there is a lack of non-absorbed GRBs at high redshift and 
a lack of heavily absorbed GRBs at low redshift. This might be the outcome of biases present in the sample.
Looking at the distribution of X--ray column densities versus redshift  a lack of non-absorbed GRBs at high redshift and 
a lack of heavily absorbed GRBs at low redshift were found in previous studies (Campana et al. 2010). 
The former might be explained in terms of more compact and dense star formation
regions in the young Universe (or to a sizable contribution from intervening systems). The latter might be interpreted as due to 
a change in the dust properties with redshift, with GRBs at redshift $z\lsim 2-3$ having a higher dust to gas ratio for 
the same X--ray column density (e.g. different grain size or composition). This will naturally provide a lack of heavily (X--ray) absorbed GRBs at small redshifts.

In the optical the presence of a large amount of absorbing material  is much less clear, since a large number of GRB afterglows are not affected 
by absorption (Kann, Klose \& Zeh 2006; Schady et al. 2010; Zafar et al. 2011; but see Greiner et al. 2011 and Covino et al. 2011, in preparation).
In this respect photoionisation of dust grains can play an important role (Lazzati, Perna \& Ghisellini 2001; Lazzati, Covino \& Ghisellini; 
Draine \& Hao 2002; Campana et al. 2007).
Moreover, the absorbing column densities measured in the optical 
based on damped Lyman-$\alpha$ absorption are a factor of $\sim 10$ lower than those measured in the X--ray band 
(Campana et al. 2010; Fynbo et al. 2009). This has been interpreted as due to photoionization of the surrounding medium 
by GRB photons (Campana et al. 2006, 2007; Watson et al. 2007; Campana et al. 2010; Schady et al. 2011).

The presence of a large amount of material is also testified by the existence of `dark' GRBs. There are several definitions of dark GRBs.
The easiest is that they do not show an optical counterpart (Fynbo et al. 2001). Since this definition is clearly related to the sensitivity 
(and availability) of the instruments used for the follow-up a more general definition is needed.
Based on the predictions of the fireball model (M\'esz\'aros \& Rees 1997) one can require that the optical to
X--ray spectral index $\beta_{OX}$ (i.e. the slope between the fluxes in the
$R$-band and at 3 keV at 11 hr after the burst) should be lower than 0.5 (Jakobsson et al. 2004). 
This will individuate optically sub-luminous bursts, i.e. fainter than expected from the fireball model.
Alternatively, with the advent of {\it Swift}, X--ray spectral slopes were more easily available and a somewhat different definition
was put forward by van der Horst et al. (2009) for which $\beta_{OX}$ is shallower than $\beta_X - 0.5$.

The darkness of a GRB can have different causes: it can be due to intrinsically optically faint GRBs, it can be due to absorption 
by intervening material within the host galaxy or it can be due to high redshift GRBs, thus being absorbed by the intergalactic medium.
Several works have addressed this topic in the {\it Swift} era when a number of facilities allowed a quick follow-up of the 
afterglows. The fraction of dark bursts has been estimated to be $\sim 25-50\%$ according to Jakobsson's definition 
(Melandri et al. 2008; Roming et al. 2009; Cenko et al. 2009; Greiner et al. 2011; Melandri et al. 2011). It is now believed that the faint 
optical afterglow emission of dark bursts might be due to a moderate intrinsic extinction at moderate redshifts. 
Greiner et al. (2011) estimated a $\sim 20\%$ contribution from high redshift ($z\gsim 4-5$) GRBs to the dark population only.

Salvaterra et al. (2011, see also Nava et al. 2011) selected a complete sample of bright GRBs based on optical observability (Jakobsson et al. 2006) and
{\it Swift} BAT peak flux $P\ge 2.6$ ph s$^{-1}$ cm$^{-2}$. The sample consists of 58 GRBs and it is complete in spectroscopic redshift 
at $90\%$ (with $95\%$ of GRBs having some constraints on the redshift).
This sample offers the opportunity to study in an unbiased way the distribution of the X--ray column densities and 
its relation to GRB darkness.

The paper is organised as follows. In section 2 we derive the X--ray absorbing column densities for the Salvaterra's sample 
and briefly describe how the slope $\beta_{OX}$ has been computed for each GRB of the sample. 
In section 3 we discuss our findings and in section 4 we draw our conclusions.

\section{X--ray absorbing column densities and spectral slope between optical and X--ray bands}

The intrinsic column densities were computed using the automated data products provided by
the {\it Swift}/XRT GRB spectra repository (Evans et al. 2009). The archive has been recently updated by 
reprocessing all the on-line GRB data products using the latest software and calibration (Evans 2011).
Therefore some of these values overwrite those reported in Campana et al. (2010). 
In Table 1 we list the column density value at the host galaxy redshift $N_H(z)$. 
These are obtained fitting an absorbed power law model to the data 
in the specified time interval when there are no strong spectral variations. 
The absorption component is modeled with {\tt PHABS} within {\tt XSPEC}. 
We consider two components one Galactic (held fixed) and a component at the redshift of the GRB ({\tt ZPHABS}).
The Galactic column density for each burst is provided by the Leiden/Argentine/Bonn (LAB) Survey of Galactic HI 
(Kalberla et al. 2005).
For those GRBs without redshift, we fix the redshift of the free absorption component to zero,
so that the resulting value provides a lower limit to the intrinsic column density. 

In the next sections we will also use the $\beta_{OX}$ index. This index is computed as the spectral index
connecting the $R$-band flux and the (unabsorbed) 3 keV flux at 11 hr from the trigger. The collection of indexes and limits for the 
burst in our sample can be found in Melandri et al. (2011). 

\begin{figure}
\begin{center}
\includegraphics[width=8.5cm]{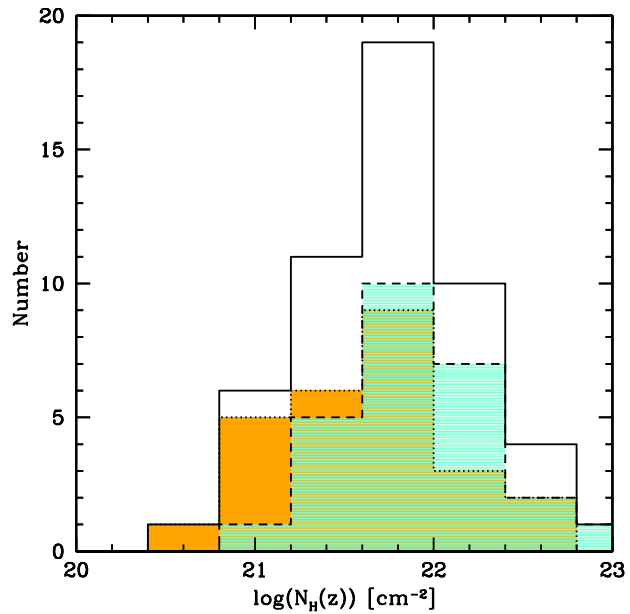}
\end{center}
\caption{Column density ($N_H$) as derived from the X--ray data of the GRBs in our complete sample. The dotted
(orange) histogram shows the distribution for GRBs with $z<1.7$ and the dashed (light blue) histogram shows the distribution for 
GRBs with $z>1.7$.}
\label{metal}
\vskip -0.1truecm
\end{figure}

\begin{table*}
\caption{Column densities for bright {\it Swift} GRBs in the Salvaterra et al. (2011) sample.}
\footnotesize{
\begin{tabular}{ccccccc}
GRB         & $z$   & $N_H(z)$                     &$\Gamma$       & $N_H(\rm Gal)$      & Time interv.    & Comments\\
                 &            & ($10^{21}\cmdue$)    &                           &($10^{20}\cmdue$) &     (s)                & (exp. time ks)\\
\hline
050318    &1.44   &$0.7^{+0.6}_{-0.5}$     &$1.97\pm0.07$&    1.9          & $3000-7\times 10^4$ & PC (23.5) \\
050401    &2.90   &$18.2^{+2.3}_{-2.3}$   &$1.84\pm0.05$&    4.4         &  $200-2000$                 & WT (1.4) \\
050416A  &0.65  &$6.4^{+0.9}_{-0.5}$     &$2.12\pm0.06$&    2.4          &  $400-3\times 10^5$   & PC (92.3) \\  
050525A  &0.61  &$2.0^{+0.9}_{-0.9}$     &$2.12\pm0.16$&    9.1          &  $6000-3\times10^5$  & PC (18.7) \\
050802     &1.71  &$1.6^{+1.6}_{-1.5}$     &$1.86\pm0.12$&    1.9           & $400-2000$                 & PC (1.5)\\
050922C  &2.20  &$3.5^{+2.2}_{-2.1}$     &$2.24\pm0.11$&    5.4          &  $400-3\times 10^5$   & PC (7.3)\\
060206     &4.05  &$14.2^{+16.0}_{-9.8}$ &$2.04\pm0.28$&    0.9          &  $700-10^5$                 & PC (4.5) \\
060210     &3.91  &$24.6^{+2.9}_{-3.5}$   &$2.09\pm0.05$&    6.1          &  $3000-10^6$               & PC (76.8) \\
060306     &3.50  &$97^{+47}_{-36}$        &$1.86\pm0.26$&   3.4          &   $300-925$                   & PC (0.6) \\
060614     &0.13  &$0.33^{ +0.18}_{-0.13}$&$1.90\pm0.06$& 1.9          & $4400-10^6$                 & PC (28.4)\\
060814     &1.92  &$20.9^{+2.3}_{-2.2}$    &$1.94\pm0.07$&   2.3          &  $250-500$                     &WT (0.2)\\   
060904A  &--       &$>2.07$                           &$3.53\pm0.43$&   1.3           & $185-225$                     & WT (0.1)\\
060908     &1.88  &$6.2^{+2.8}_{-2.5}$      &$2.12\pm0.18$&    2.3          &  $200-10^5$                  & PC (11.6)  \\
060912A  &0.94  &$3.2^{+1.5}_{-1.3}$      &$2.02\pm0.18$&    3.9           &  $200-2000$                  & PC (1.7)\\
060927     &5.47  & $<36$                            &$1.94\pm0.16$&    4.6          &  $100-10^4$                  & PC (3.5) \\
061007     &1.26  &$5.1^{+0.3}_{-0.3}$      &$1.85\pm0.02$&    1.8          &  $90-2000$                    &  WT  (1.9)\\
061021     &0.35   &$0.73^{+0.2}_{-0.1}$   &$1.99\pm0.03$&    4.5          &  $3000-3\times10^5$   & PC (83.1)\\
061121     &1.31  &$5.4^{+0.8}_{-0.5}$      &$1.88\pm0.05$&    4.0           &  $600-3\times10^5$     & PC (43.1) \\
061222A  &2.09  &$44.8^{+5.4}_{-3.0}$    &$2.11\pm0.06$&    9.0           &  $3\times 10^4-2\times 10^5$ & PC (49.5)\\
070306     &1.50  &$26.8^{+4.7}_{-4.3}$   &$1.88\pm0.12$&    2.9           &  $10^4-4\times10^4$    & PC (6.0)\\
070328     &{\bf --}  &$2.4^{+0.2}_{-0.2}$      &$2.24\pm0.05$&    2.6           &  $350-1000$                  & WT (0.7)\\
070521     &1.35   &$54^{+13}_{-11}$       &$1.78\pm0.20$&    2.9           & $3000-10^4$                 & PC (1.9)\\
071020     &2.15   &$6.8^{+1.7}_{-1.6}$     &$1.87\pm$0.07&    5.1           &  $70-300$                       & WT (0.2)\\
071112C  &0.82   &$1.4^{+0.5}_{-0.5}$     &$1.82\pm0.08$&    7.4            &  $90-280$                      & WT (0.2)\\
071117     &1.33    &$10.9^{+2.1}_{-3.1}$   &$2.05\pm0.18$&    2.3          &  $2900-6.2\times 10^4$& PC (19.0)\\
080319B  &0.94   &$1.7^{+0.1}_{-0.1}$      &$1.78\pm0.02$&    1.1           &  $800-2000$                   &  WT (1.7) \\
080319C  &1.95   &$5.5^{+2.5}_{-2.3}$      &$1.61\pm0.10$&    2.2           &  $200-3\times10^5$      & PC (4.1) \\
080413B  &1.10   &$1.9^{+0.6}_{-0.4}$      &$1.97\pm0.07$&    3.1           &$6\times 10^3-10^6$     &PC (40.6)\\
080430    &0.77    &$3.5^{+0.8}_{-0.6}$       &$2.03\pm0.10$&    1.0           &  $5000-3\times10^5$  & PC (10.9)\\
080602A &1.40    &$6.7^{+2.4}_{-2.1}$      &$2.01\pm0.15$&    3.5           & $200-800$                      & PC (0.6)\\     
080603B  &2.69   &$7.3^{+2.9}_{-2.7}$      &$1.84\pm0.10$&    1.2            &  $100-250$                    & WT (0.2)\\
080605    &1.64    &$9.0^{+0.9}_{-0.9}$      &$1.76\pm0.04$&    6.7            &  $100-800$                   & WT (0.6)\\
080607    &3.04    &$22.8^{+5.7}_{-4.2}$    &$2.14\pm0.10$&    1.7            &  $4000-6\times10^4$  & PC (19.7)\\
080613B  & --       &$>0.5$                             &$1.31\pm0.12$&    3.2            &$105-190$                      & WT (0.1)\\   
080721    &2.59   &$10.4^{+0.6}_{-0.6}$     &$1.81\pm0.02$&    6.9           &  $100-2000$                  & WT (1.3)\\
080804    &2.20   &$1.4^{+1.9}_{-1.1}$        &$1.82\pm0.09$&    1.6           &  $200-3\times10^5$     & PC (12.6)\\
080916A  &0.69   &$8.0^{+3.2}_{-1.9}$       &$2.26\pm0.15$&    1.8           &  $2\times10^4-10^6$    & PC (172.9)\\
081007    &0.53   &$4.8^{+0.9}_{-1.2}$        &$2.04\pm0.18$&    1.4           &  $5000-4\times10^5$   & PC (9.7)\\
081121     &2.51   &$1.9^{+1.6}_{-1.5}$        &$1.93\pm0.06$&     4.0           &  $3000- 2\times10^6$ & PC (36.3)\\
081203A  &2.05   &$4.5^{+1.1}_{-1.0}$       &$1.74\pm0.05$&     1.7           &  $200-600$                    &  WT (0.4)\\
081221   &2.26    &$26.1^{+3.8}_{-3.6}$      &$2.04\pm0.09$&     2.0           &  $300-500$                    & WT (0.2)\\      
081222   &2.77     &$6.0^{+1.1}_{-1.0}$       &$1.96\pm0.04$&     2.2           &  $60-1000$                    &  WT (0.8)\\
090102   &1.55   &$5.0^{+2.5}_{-2.2}$         &$1.73\pm0.13$&    4.1            &  $700-7\times10^4$     &PC (1.8)\\
090201   & $<4$ &$>3.85$                             &$2.01\pm0.16$&    4.9            &$3000-6000$                  & PC (1.7) \\  
090424   &0.54   &$4.1^{+0.6}_{-0.5}$         &$1.94\pm0.08$&    1.9            &  $2000-3\times10^6$   & PC (14.9)\\
090709A &$<3.5$&$>1.82$                           &$2.04\pm0.13$&     6.6           &  $4000-1.5\times10^4$ & PC (4.4) \\  
090715B &3.00    &$7.6^{+2.5}_{ -2.8}$      &$2.01\pm0.10$&    1.3            & $4000-10^5$                  & PC (28.8)\\
090812    &2.45    &$12.0^{+7.2}_{-6.6}$     &$2.10\pm0.23$&     2.3           &  $10^4-7\times10^4$     & PC (7.7) \\
090926B  &1.24   &$13.9^{+1.6}_{-1.5}$     &$1.97\pm0.08$&   1.9              &  $130-300$                   & WT (0.2)\\
091018     &0.97   &$1.0^{+0.9}_{-0.8}$     &$1.84\pm0.12$&      2.8             &  $150-1000$                & PC (0.9)\\
091020     &1.71   &$5.8^{+1.7}_{-1.6}$       &$1.82\pm0.10$&    1.4             &  $200-400$                   & WT (0.2)\\
091127     &0.49   &$0.76^{+0.35}_{-0.5}$   &$1.80\pm0.11$&      2.8           & $6000-2\times10^4$ & PC (2.0)\\
091208B  &1.06   &$8.3^{+4.3}_{-3.4}$       &$2.16\pm0.27$&   4.9              & $200-600$                  & PC (0.4)\\
100615A  &--         &$>10.1$                           &$2.43\pm0.32$&    3.3             & $200-2000$                & PC (1.4)\\
100621A  &0.54    &$18.0^{+1.2}_{-1.1}$    &$2.86\pm0.09$&      2.9           & $130-200$                  & WT (0.1)\\
100728B  &2.11    &$4.3^{+3.1}_{-2.5}$       &$2.08\pm0.18$&      6.2          &  $4000-3\times 10^4$ & PC (8.5)\\
110205A  &2.22    &$3.5^{+1.9}_{-1.4}$       &$2.11\pm0.09$&      1.6          & $5000-5\times 10^4$  & PC (18.8)\\
110503A  &1.61    &$3.53^{+0.67}_{-0.64}$ &$1.80\pm0.05$&      2.6         & $200-700$                 & WT (0.5)\\
\hline
\end{tabular}
}

\noindent Errors and upper limits are at $90\%$ confidence level obtained with a $\Delta \chi^2=2.71$.
\end{table*}

\section{Discussion}

\subsection{Distribution of absorbing column densities}

The distribution of the X--ray column densities is shown in Fig. 1. The overall distribution can be well described by a Gaussian
with (logarithmic) mean 21.7 and standard deviation 0.5 (the median of the distribution is very close to the mean).
This is consistent with the column density distribution obtained
considering all GRBs with known redshift (Campana et al. 2010). The distribution of column densities as a function of redshift 
is shown in Fig. 2. It is apparent from Fig. 2 that there is a trend of increasing column densities with redshift. However, the lack of 
mildly absorbed GRBs at high redshift is not statistically overwhelming. To see if there is a real difference we 
cut the sample at the mean redshift ($z=1.7$), and make a Kolmogorov-Smirnov (KS) test on the two distributions. 
We obtain a probability of {\bf 9}$\%$ that the two distributions come from the same
parent population. No firm conclusions can therefore be drawn. 
A cut at $z=1$ results in a {\bf 0.5}$\%$ KS probability. This might indicate a difference in the intrinsic absorption column densities 
(see below).
With respect to a non-complete sample (Campana et al. 2010),  we note that the high-column density region at low redshift is 
here more populated. This indicates a bias present in the non-complete sample:
it is difficult to obtain the redshift of highly-absorbed low-redshift GRBs (likely due to a higher optical absorption).

\begin{figure}
\begin{center}
\includegraphics[width=8.5cm]{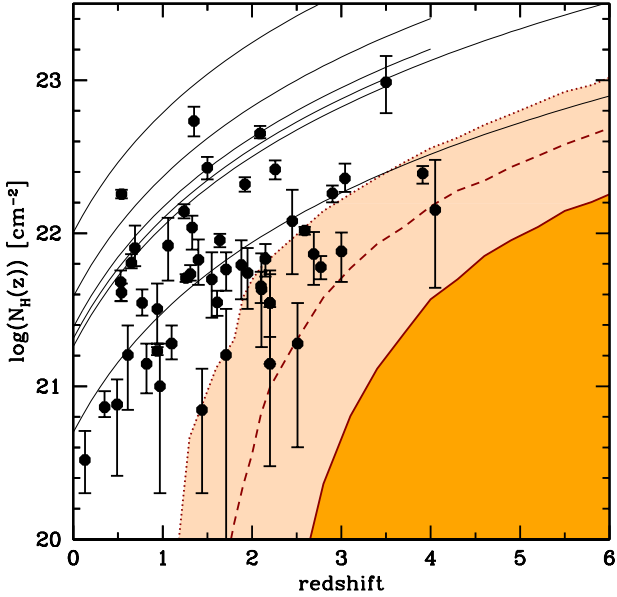}
\end{center}
\caption{Column density distribution versus redshift. Error bars have been computed from X--ray spectral fitting and 
are at $90\%$ confidence level. Upper limits are at $90\%$ confidence level. Light grey lines represent the possible column density values
for GRBs without redshift, scaling the $z=0$ value with $(1+z)^{2.6}$ within the range of allowed redshifts (Galama \& Wijers 2001).
Darker (orange) region limited by a continuous line marks the mean contribution in the observed column density $N_H(z)$ 
resulting from intervening systems as a function of redshift. The lighter (light orange) region limited by a dotted line 
marks the maximal $90\%$ line of sight contribution of intervening systems to the observed column density. 
The dashed line marks the mean contribution to $N_H(z)$ in the case of doubling the population of
intervening systems as suggested by the comparison of quasars and GRB studies of intervening systems (see text, section 3.1).}
\label{metal}
\end{figure}

The lack of mildly-absorbed high redshift GRBs remains. This has been interpreted as due to the presence of
absorbing matter along the line of sight not related to the GRB host galaxy. This can be either diffuse (i.e. located in diffuse structures like the 
filaments of the Warm-Hot Intergalactic medium - WHIM,  Behar et al. 2011) or concentrated into intervening systems (i.e. galaxies or clouds along
the line of sight, Campana et al. 2010). 
Based on quasar studies (Wolfe et al. 2005; P\'eroux et al. 2003; Noterdaeme et al. 2009; Prochaska, O'Meare \& Worseck 2010),
we can evaluate the contribution of the intervening systems 
to the observed column density by simulating their distribution. 
In particular, we assumed a number distribution of intervening systems, based on damped (and sub-damped) Lyman-$\alpha$ systems, 
with a redshift dependence $n(z)\propto (1+z)^{0.26}$ for $z\le2.3$ 
and $n(z)\propto (1+z)^{2.78}$ for $z> 2.3$. For the contribution of the intervening systems in terms of absorption we adopted 
a threefold equation (measuring column densities in cm$^{-2}$ units):
for $\log{N_H}<18.2$ we consider $n(\log{N_H})\propto\log{N_H}^{-1.9}$, 
for $18.2\le \log{N_H}\le20.6$ $n(\log{N_H})\propto \log{N_H}^{-0.8}$, 
and for $\log{N_H}>20.6$ $n(\log{N_H})\propto\log{N_H}^{-1.4}$ 
(Wolfe et al. 2005; P\'eroux et al. 2003).
The contribution of each intervening system is weighted considering the system as if it was at the redshift of the GRB, i.e. weighting its
intrinsic column density as $((1+z_{GRB})/(1+z_{DLA}))^{2.6}$ (where $z_{GRB}$ is the redshift of the GRB and $z_{DLA}$ the redshift of the intervening system). 
We set up a MonteCarlo simulation considering systems in the $\log(N_H/{\rm cm^{-2}})=17.2-22$ range and probing 10,000 lines-of-sight.
Because the GRB column densities were calculated assuming solar metallicity, for comparison we also assumed solar metallicities 
for the evaluation of the contribution of the intervening systems. It is important to note that GRB hosts have typically sub-solar metallicities, 
in which case the assumption of solar metallicity would lead to the equivalent column density being underestimated. 
All data points in Fig. 2 can therefore be (usually) considered as being lower-limits on $N_H$.
The $90\%$ confidence envelope of the simulated line-of-sights as a function of redshift is shown in Fig. 2 (dashed line). 
The average contribution is also shown (continuous line and dark orange region).
These calculations clearly provide just an indicative estimate and are subject to uncertainties related to
modeling the number density evolution in redshift of these systems (e.g. Ribaudo, Lehner \& Howk 2011).
It is apparent from Fig. 2 that a GRB lying along the `average' line of sight experiences a too low increase of the 
intrinsic column density due to intervening systems with respect to the observed increase of GRB column densities with redshift 
(even if a less favorable line of sight might fully account for the observed increase at high redshifts).

Studies on strong intervening systems in quasars and GRB spectra have shown a larger occurrance of intervening systems in the latter lines of sight 
(e.g. Prochter et al. 2006). These systems are mainly identified through strong, rest-frame equivalent width (EW) $> 1$ \AA\ Mg II 
$\lambda\lambda$ 2796, 2803 absorption lines. In a study with a large sample of GRBs, Vergani et al. (2009) 
confirm the presence of this effect and set the discrepancy to a factor of $\sim 2$. 
Even if the reason for this discrepancy is still not fully understood, Budezynski \& Hewett (2011) show that 
this discrepancy is likely related to a lack of quasars heavily absorbed along their line of sight.
Given this observational result, we artificially increased the number of intervening systems based on quasar studies by a factor of 2.
The resulting mean contribution derived from the intervening systems is shown in Fig. 2 with a dashed line. This line follows nicely the 
increase of the intrinsic column density with redshift providing a plausible explanation for this effect.

\begin{figure}
\begin{center}
\includegraphics[width=9cm]{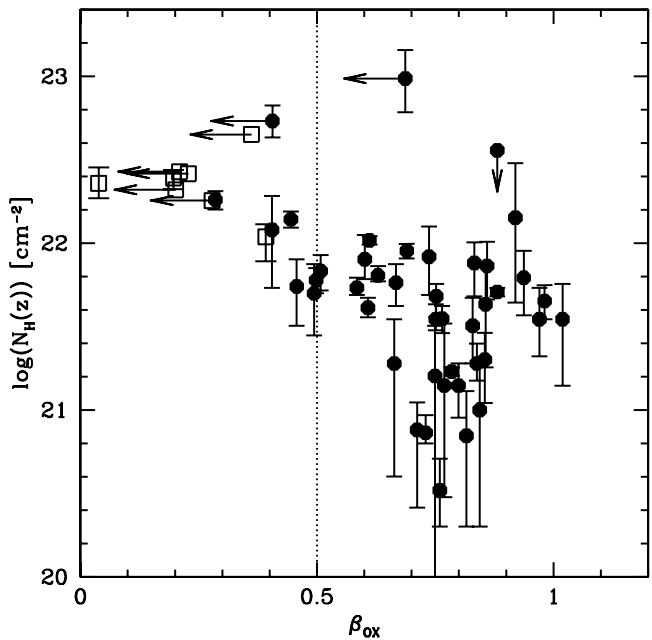}
\end{center}
\caption{Column density ($N_H$) as a function of the spectral index $\beta_{OX}$ (Melandri et al. 2011) for the GRBs in our 
complete sample. Limits on the column density and $\beta_{OX}$ are indicated with arrows. 
The dashed line for $\beta_{OX}=0.5$ divides dark and non-dark GRBs according to Jakobsson et al. (2004). 
Open squares (filled circles) indicate dark (non-dark) GRBs according to van der Horst et al. (2009).
GRBs with lower limits on the column density (i.e. without redshift information) are not shown.}
\label{metal}
\end{figure}

\subsection{Relation between X--ray absorption and GRB `darkness'}

Given our complete sample of bright GRBs we investigate the connection between the X--ray absorption and the GRB darkness. 
The GRB darkness can be caused by several effects that can be divided into two main classes: $i)$ intrinsic, i.e. due
to some physical mechanism hampering the optical emission or $ii)$ environmental, i.e. due to intrinsic absorption and/or to high redshift.
Considering the $\beta_{OX}$ values computed in Melandri et al. (2011), there are 19 GRBs in our complete sample that can be 
classified as dark according to Jakobsson et al. (2004) and 12 according to van der Horst et al. (2009). 
Out of them, 4 (common to both definitions) do not have any redshift information. 
In Fig. 3 we show the distribution of $\beta_{OX}$ as a function of the intrinsic column density for the GRBs of our complete sample
(we did not include the few GRBs without a redshift determination). 
It is apparent that for bursts with $\beta_{OX}<0.5$ (i.e. dark according to the Jakobsson's definition) all but three (all with $0.45<\beta_{OX}=0.5$)
have an intrinsic column density larger than $\log(N_H/{\rm cm^{-2}})\gsim 22$. This same $N_H$ limit is valid for all the bursts that are dark 
according to the van der Horst's definition (Fig. 3).

Comparing  the intrinsic column density distribution of dark and non-dark GRBs (taking only the ones with known redshift, i.e. 15--38 and 8--45 for the 
Jakobsson's and van der Horst's definition, respectively) we obtain a KS probability 
of $2\times 10^{-6}$ ($4.8\,\sigma$) for the Jakobsson's definition and  $1\times 10^{-5}$ ($4.4\,\sigma$) for the van der Horst's definition 
(the lower value is due to the smaller number of dark GRBs).
We also note that if the 4 GRBs classified as dark and without redshift information would have a redshift in line with the mean of the sample, 
then they would have an intrinsic column density $N_H(z)\gsim 10^{22}$ cm$^{-2}$.
These results indicate that the intrinsic absorption as evaluated in the X--ray band is highly correlated with the darkness of a GRB. 

\section{Conclusions}

Salvaterra et al. (2011) selected a complete sample of bright GRBs with a high degree of completeness in redshift.
In a series of papers we investigate the impact of this sample on GRB studies. Here we focus on the properties of the sample
with respect to the intrinsic X--ray absorption. 
We found that the intrinsic column density distribution of our complete sample is consistent with the total distribution of column 
densities presented in Campana et al. (2010). The mean of the two distributions are in fact $21.7\pm0.5$ and $21.9\pm0.1$, 
respectively. 
This likely indicates that the GRB brightness, as well as any other bias present in the total sample of GRBs with redshift (e.g. dust), does
not heavily influence the total distribution of intrinsic column densities. 

At variance with the total distribution presented in Campana et al. (2010), we see in the complete sample presented here 
that the region at high column densities and low redshift is now more populated by GRBs. This clearly reveals a bias present in the non-complete
sample, where this region is heavily underpopulated due to the lack of  a redshift determination of dark bursts.

Even if not statistically compelling there is an increase of the intrinsic column density with redshift  (this is more apparent in the 
full sample of GRBs with redshift, Campana et al. 2010). We evaluate the mean contribution to $N_H(z)$ due to the intervening systems
along the GRB line of sight. We find that, if we take into account the larger number of observed systems affecting the line of 
sight of GRBs with respect to the quasar one (Vergani et al. 2009), the population of 
sub-Damped Lyman-$\alpha$ and Damped Lyman-$\alpha$ systems can account for the increase with redshift of $N_H(z)$.
It would be interesting to confirm this directly through the study of high-$z$ GRB lines of sight. 
Unfortunately this effect plays a significant role at very high redshift, where the number of GRB afterglow spectra is very low. 
It is indeed difficult to measure absorption from Lyman-$\alpha$. However, the column density of neutral gas can still be traced by weakly ionised metal lines 
(e.g. Zn II, Si II), which in fact is a more logical method of comparing absorption in X--rays and the optical, given that the X--rays are absorbed by metals and 
not neutral hydrogen (e.g. Schady et al. 2011).
%Moreover,  it is difficult to determine directly the $N_{H}$ of the intervening absorbers from the optical afterglow spectra because of the 
%increasing number of systems for which the Lyman-$\alpha$ transition is not detected due to the HI intergalactic medium absorption 
%at $z<z_{GRB}$. 
%In addition, the MgII doublet transition used to identify these systems is outside the optical range from $z\gsim2.5$. 
The X-shooter instrument mounted at the ESO/VLT offers the best opportunities for these studies.

Making use of the $\beta_{OX}$ computed by Melandri et al. (2011), we found a strong correlation between GRB darkness 
and X--ray absorbing column densities. Since metals are a key ingredient for dust production (Draine 2003), our findings are consistent
with a picture in which the darkness of a GRB is in most cases due to absorption by circumburst material.

\section{Acknowledgments}
SC thanks Darach Watson and Phil Evans for useful conversations. This work has been supported by ASI grant I/004/11/0.
This work made use of data supplied by the UK {\it Swift} Science Data Centre at the University of Leicester.

{}

\end{document}